\documentclass{article}
\DeclareUnicodeCharacter{0306}{}

\usepackage[utf8]{inputenc}
\usepackage[T1]{fontenc}
\usepackage[style=apa, backend=biber, sorting=ynt]{biblatex}
\usepackage[acronym]{glossaries}
\usepackage{graphicx}
\usepackage{siunitx}
\usepackage{geometry}
\usepackage{authblk}
\usepackage[font=small]{caption}
\usepackage{enumitem}

\makeglossaries

\begin{document}

\newacronym{spc}{SpC}{spinal cord}
\newacronym{bst}{BSt}{brainstem}
\newacronym{hth}{HTh}{hypothalamus}
\newacronym{ctx}{CTX}{cortex}
\newacronym{bg}{BG}{basal ganglia}
\newacronym{cb}{CB}{cerebellum}
\newacronym{m1}{M1}{primary motor cortex}
\newacronym{sma}{SMA}{supplementary motor area}
\newacronym{gpe}{GPe}{external segment of the globus pallidus}
\newacronym{gpi}{GPi}{internal segment of the globus pallidus}
\newacronym{vta}{VTA}{ventral tegmental area}
\newacronym{snc}{SNc}{substantia nigra pars compacta}
\newacronym{snr}{SNr}{substantia nigra pars reticulata}
\newacronym{sc}{SC}{superior colliculus}
\newacronym{mlr}{MLR}{midbrain locomotor region}
\newacronym{mrf}{MRF}{medullary reticular formation}
\newacronym{scn}{SCN}{suprachiasmatic nucleus}
\newacronym{mf}{MF}{mossy fiber}
\newacronym{gc}{GC}{granule cell}
\newacronym{pc}{PC}{Purkinje cell}
\newacronym{cf}{CF}{climbing fiber}
\newacronym{mn}{MN}{motor neuron}
\newacronym{cpg}{CPG}{central pattern generator}
\newacronym{cff}{CFF}{convergent force field}
\newacronym{da}{DA}{dopamine}
\newacronym{ttfl}{TTFL}{transcription-translation feedback loops}
\newacronym{rl}{RL}{reinforcement learning}
\newacronym{td}{TD}{temporal difference}
\newacronym{rpe}{RPE}{reward prediction error}
\newacronym{rnn}{RNN}{recurrent neural network}

\renewcommand{\thefootnote}{\arabic{footnote}}

\title{Time, control, and the nervous system\footnotemark[2]
\footnotetext[2]{After \textit{Time, communication, and the nervous system}, Norbert Wiener---(1948)
}}

\author[1,*]{Caroline Haimerl}
\author[1,*]{Filipe S. Rodrigues}
\author[1]{Joseph J. Paton}

\affil[1]{
    Champalimaud Centre for the Unknown, Champalimaud Neuroscience Program}
\affil[*]{These authors contributed equally to this article}

\date{}

\maketitle

\begin{abstract}
    Because organisms are able to sense its passage, it is perhaps tempting to treat time as a sensory modality, akin to vision or audition. Indeed, certain features of sensory estimation, such as Weber’s law, apply to timing and sensation alike~\parencite{Gibbon1977-wn, Pardo-Vazquez2019-ea}. However, from an organismal perspective, time is a derived feature of other signals, not a stimulus that can be readily transduced by sensory receptors. Its importance for biology lies in the fact that the physical world comprises a complex dynamical system. The multiscale spatiotemporal structure of sensory and internally generated signals within an organism is the informational fabric underlying its ability to control behavior. Viewed this way, temporal computations assume a more fundamental role than is implied by treating time as just another element of the experienced world~\parencite{Paton2018-nn}. Thus, in this review we focus on temporal processing as a means of approaching the more general problem of how the nervous system produces adaptive behavior.
\end{abstract}

\providecommand{\keywords}[1]
{
  \small	
  \textbf{\textit{Keywords---}} #1
}
\vspace{.5cm}
\keywords{timing, control theory, basal ganglia, cerebellum, cerebral cortex, behavior}

\vfill

Corresponding authors:\\
\indent\indent\texttt{caroline.haimerl@research.fchampalimaud.org}\\
\indent\indent\texttt{filipe.rodrigues@neuro.fchampalimaud.org}\\
\indent\indent\texttt{joe.paton@neuro.fchampalimaud.org}

\vspace{.5cm}

\newpage

\section{Introduction}
\label{sec:introduction}
Much of biology can be viewed through the lens of control. Our prokaryotic forbearers, for example, needed to control osmotic balance. Calcium entering a cell through a tear in its membrane is thought to have triggered two processes: (\textit{a}) contraction of a ring of actomyosin to pull the edges of the tear together, and (\textit{b}) exocytosis of vesicles to add new membrane required to patch it~\parencite{Brunet2016-td}. The coupling of a damage signal, rising calcium levels, to the repair response represents a simple form of feedback control, fundamentally reactive in nature. However, if the cell could somehow anticipate membrane damage, it might preempt it. The evolution of mechanosensitive calcium channels that open in response to membrane tension appears to have enabled such anticipatory ability~\parencite{Martinac1990-re}. Membrane stress opens these channels, producing calcium influx, translating what was a sign of membrane damage to an earlier point in time. This can be seen as a logic system that implicitly embeds the prediction that membrane tension leads to membrane damage, hardcoded into the genome by virtue of the stability of the temporal relationships that it exploits. We highlight this example to emphasize that the benefits of predictive control extend beyond nervous systems, representing a general principle in biology~\parencite{Baluska2016-qe}. Most biological systems are composed of layer upon layer of such mechanisms, involving individual cells, nervous systems, or organisms as a whole. 

Controllers couple observations to commands to achieve objectives (Figure \ref{fig:1}A). Nervous systems control behavior through interaction with a variety of sensor and effector systems, and multiple lines of research have sought to identify control objectives as a normative means of explaining aspects of behavior, including continuous motor control~\parencite{Wolpert1995-qv,Todorov2004-mm}, discrete action selection~\parencite{Daw2005-ry,Cisek2011-bt}, or behavior in general~\parencite{Gibson1977-vy,Friston2010-gr}. However, even if the brain as a whole is optimized for a global control objective (e.g., maximizing cumulative expected reward, minimizing free energy), it is clear that this is achieved through the concerted action of distinct interacting modules. We argue that existing computational theories for adaptive behavior either are too general to account for this modularity or focus on only specific neural circuit mechanisms and hence are too narrow to explain the breadth of computations required to account for organismal behavior as a whole. A means of tying together individual hypotheses regarding components of nervous systems into more cohesive, integrative explanations for the neural basis of behavior is currently lacking. 

\begin{figure}[htbp]
    \centering
    \includegraphics[width=1\linewidth]{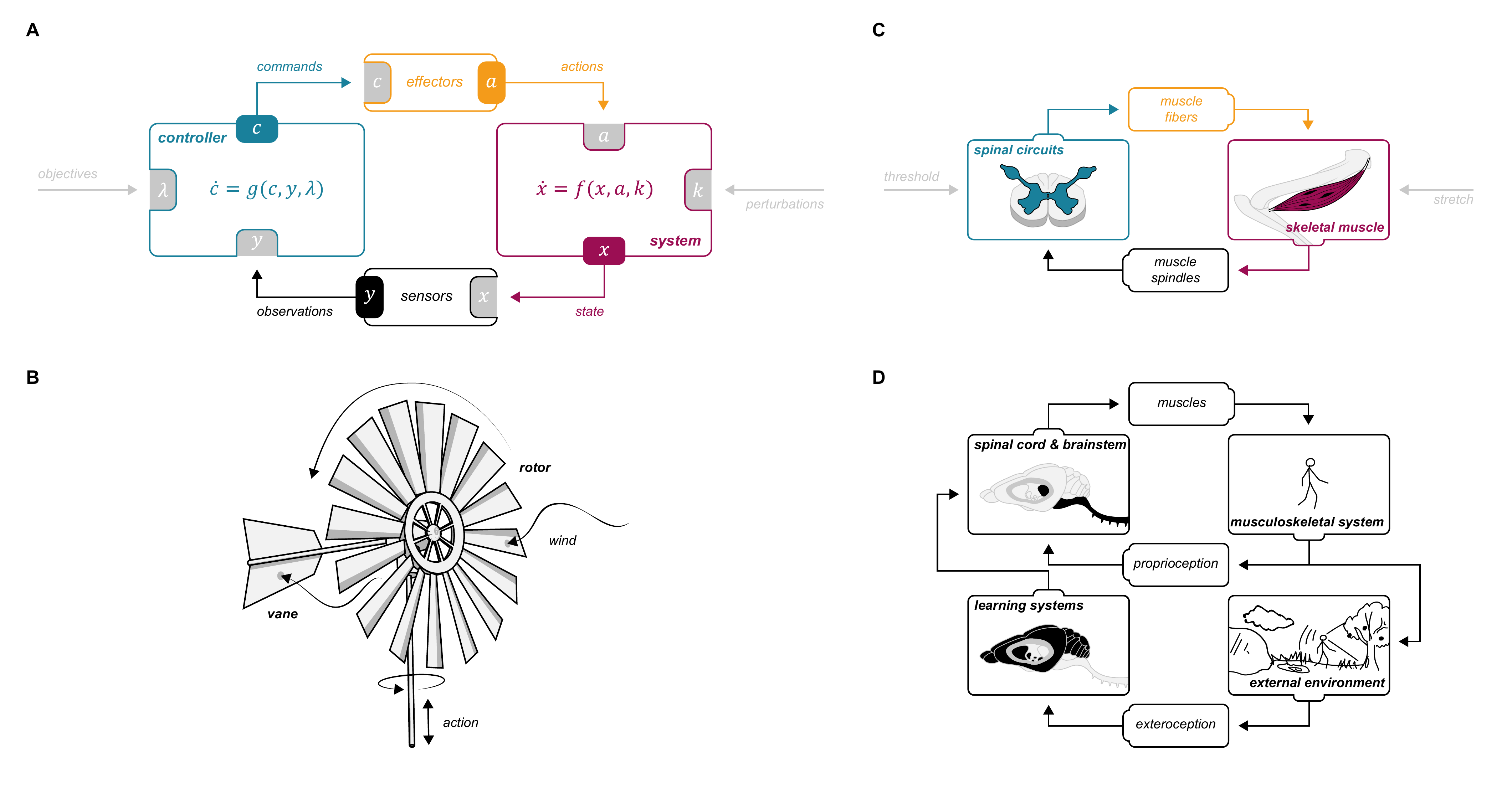}
\end{figure}
\begin{figure}[t]
    \caption{
    \textbf{A control-oriented view of modular brain architecture highlights the centrality of temporal processing for behavior.} \textbf{A)} Canonical control diagram wherein a controller emits time-varying commands that operate through effectors to produce actions. Those actions influence the temporal evolution of the state of the controlled dynamical system. That system evolves as a function $f()$ of its current state $x$, the actions of the controller, $a$, and any other external perturbations, $k$, the system is subject to. The state of the system $x$ passes through sensors to give rise to the controller’s observations, $y$, of the system. Controllers output commands according to a control law $g()$ that is a function of the controller’s current output commands, $c$, system observations, $y$, and inputs defining the current objective or target for the controller, $\lambda$. We note that controllers are fundamentally constrained by the dynamics of the system they aim to control, effectively embedding information about the temporal structure of the dynamical systems in which they operate into their control laws. \textbf{B)} A traditional windmill serves as a metaphor for logic systems. The rigid coupling of the tail vane to the rotor blades embeds knowledge of the fact that optimal harnessing of the available wind flow into useful mechanical action (e.g., pumping or grinding) is achieved by minimizing side winds hitting the vane. \textbf{C)} A control diagram for the muscle stretch reflex. In this case, the relevant state variable corresponds to muscle length, which can be influenced by either perturbations (the physician’s hammer on the patellar tendon) or actions that result from activation of spinal motor neurons (i.e., muscle contraction). Muscle length is transformed into observations the spinal reflex circuit can receive through muscle spindle organs, where they feed into the control law implemented by the reflex circuit. The parameters of this control law can be influenced by inputs, such as thresholds for activation, that ultimately define some behaviorally relevant objective for the reflex (e.g. your joint is at a particular angle so you need to activate the reflex at a particular range of muscle lengths accordingly). \textbf{D)} Ultimately, nervous systems are composed of layer upon layer of interconnected control schemes such as that described in panel C for the muscle stretch reflex. In some cases, the control laws are extremely flexible, subject to highly complex, modifiable, and experience-dependent parameterization, as is the case with learning systems, but stereotyped circuit architectures within each brain system can be seen as laying down the overall structure and optimality conditions for the control laws contained within. Depending on the brain structure, the upstream supplier of observations regarding the state of the target system might be sensory organs (proprioceptive, exteroceptive, or interoceptive), or nervous transmission from another brain circuit. On the output side, commands may influence motor neurons to control some effector relatively immediately, or provide objectives for downstream controllers, setting their parameters so as to achieve a goal state. }
    \label{fig:1}
\end{figure}

In the current review, we explore the idea that time may provide a path towards more unified views of brain function, because both the nervous system as a whole and its constituent parts are controllers that evolved to achieve target states within a dynamical system. A system’s dynamics specify how its state evolves through time, information that gets absorbed into and exploited by neural control mechanisms to better estimate and guide the system towards a relevant goal state at hand (Figure~\ref{fig:1}A). This core function of neural systems comes in many forms and scales, each specialized for the problem a particular circuit has evolved to deal with. Jointly these circuits form a layered, modular architecture that comprises the nervous system at large.

\subsection{Logic systems: strong anticipation across multiple timescales}
\label{subsec:logic-systems}
We begin our description with what we refer to as logic system controllers. Imagine a simple windmill. The two main structural components, the rotor and the tail vane, apply different logical operations to an environmental feature, wind. The vane, if subjected to wind forces in a direction normal to its face, rotates the face of the rotor about the vertical axis. This motion ceases at the position where such forces are minimized, corresponding to the optimal position for transforming linear flow of air into continuous rotational motion of the rotor and by extension, useful action (Figure~\ref{fig:1}B). Our windmill is in some sense flexible, adaptive to change in the direction of wind, but its mechanics are rigid. One might also say it embeds temporal regularities that underpin a prediction because forces normal to the face of the vane “predict” more efficient transformation of air flow into rotational energy. However, this prediction is not computed explicitly from any internal model of the system---weak anticipation---but instead arises from appropriate dynamical couplings that exist within the system itself---strong anticipation~\parencite{Dubois2003-tm,Stepp2010-qi}. We argue that many brain circuits can similarly be seen as implementing such logical, strong-anticipatory operations, achieving teleologically adaptive flexibility through the coupling of rigid mechanisms, similar to certain engineering approaches to robotic control~\parencite{Brooks1991-lj}. Note that we use logic here to contrast with learning, not as cognitive operations involving the flexible application of rules. Depending on the system and function, these hardwired predictions can span milliseconds to years, and inform more continuous or more discrete forms of control. Key examples include the \gls{spc}, \gls{bst}, and \gls{hth}, which map together current and anticipated states with actions at progressively longer timescales and varying granularity. In all three exemplars of logic-like control, circuit function is primarily dictated not by experience-dependent plasticity private to each individual, but rather by a rigid proximity to sensors (intero-, proprio- and/or exteroceptive) and effectors (muscles and/or glands), bridged by stereotypical wiring motifs that both reflect and exploit evolutionarily stable spatiotemporal regularities in the organism and its environment. 

\subsection{Spinal cord: a hierarchy of rapid, relatively hardwired behavioral primitives}
\label{subsec:spinal-cord}
The \gls{spc} serves as a highway for transmitting sensory and motor information and autonomously controlling reflexive and rhythmic behaviors. Sensory input enters dorsally, and motor commands exit ventrally, allowing rapid responses through reflex arcs. One illustrative example is the monosynaptic stretch reflex, where muscle spindles\footnote{Specialized sensory organs located within skeletal muscles that detect changes in muscle length and rate of stretch.} (sensors) convert increases in the length (state) of a skeletal muscle (system) into afferent signals (observations) that drive efferent activity (commands) that in turn causes muscle fibers (effectors) to contract (action), thus countering the initial stretch (perturbations)~\parencite{Liddell1924-pf}. This behavior involves two types of spinal motor neurons (\glsunset{mn}\glspl{mn}, controller): alpha \glspl{mn} directly activate extrafusal fibers\footnote{Contractile fibers of a muscle that make up the bulk of its mass, and are responsible for generating the force needed for movement and posture maintenance.}, while gamma \glspl{mn} adjust the responses of muscle spindles by regulating intrafusal fiber\footnote{Non-force-generating fibers around which the sensory endings of spindle afferents are coiled within muscle spindles.} tone (Figure \ref{fig:1}C). This arrangement allows the muscle to contract while preserving spindle feedback~\parencite{Hunt1951-ob} and creates two distinct control possibilities for descending commands. They can drive alpha \glspl{mn} directly with temporally patterned continuous input, or send discrete, intermittent signals to gamma \glspl{mn} to produce a similar profile of muscle shortening by biasing spindle proprioception and relying on local reflexive dynamics of the neuromuscular system as a whole to bring about movement~\parencite{Feldman1986-vx}.

Both control schemes therefore involve continuous sensory signals and effectors which allow simple mathematical operations such as online proportional control~\parencite{Ogata2010-fm} to avoid large discrete jumps that require compensation and could risk stability, while simultaneously allowing for precise, autonomous adjustments to perturbations~\parencite{Astrom2021-ul}. Spinal reflexes provide hardwired responses that present tunable mechanisms at the base of behavior~\parencite{Sherrington1910-xn}. According to this view, descending inputs, themselves outputs of other control functions, can specify the parameters of reflexes, comprising a layered arrangement of control mechanisms that hierarchically constructs the global characteristics of movement (Figure \ref{fig:1}D)~\parencite{Bernstein1967-zf,Feldman1986-vx,Latash2007-ne}.

Ascending in complexity from the stretch reflex, the flexion reflex retracts an appendage from harm and relies on mutual inhibition within the \gls{spc} to coordinate opposing flexor and extensor muscles groups across joints and limbs~\parencite{Sherrington1905-xh,Renshaw1941-zc,Jankowska1972-xn,Tripodi2011-wn}. Such composite reflexes can be elicited in spinalized frogs~\parencite{Giszter1993-fj}, rats~\parencite{Schouenborg1992-zs}, cats and dogs~\parencite{Sherrington1910-xn}, indicating a highly conserved function for spinal circuits in simplifying the task for higher levels in a hierarchy for control. In a more elaborate example, microstimulation of spinal interneurons elicits a force vector measured at the limb. As limb position is varied, and stimulation repeated, a \gls{cff} over limb locations can be constructed. Force fields of one stimulation site tend to point towards a location in limb space, and stimulation at multiple sites produces approximate linear combinations of site-specific \glspl{cff}, a sign of compositionality~\parencite{Bizzi1991-du}. From the perspective of other local and/or upstream circuits, this invariance to initial conditions may greatly simplify the control problem because only the movement target requires specification, and not the specific pattern of muscle activations required to acquire it. Chaining together similar motor synergies~\parencite{Latash2007-ne,DAvella2003-go} through time, and reflexively coordinating them across limbs, in principle allows spinal \glspl{cpg} to generate rhythmic movements that characterize vertebrate locomotion~\parencite{Grillner1985-lz,Ijspeert2013-yd}. 

Altogether, the \gls{spc} (Figure \ref{fig:2}A-C) provides the brain with configurable, autonomous control policies, forming a basis for more flexible behavior. Each highlighted example of \gls{spc} circuit function---a muscle stretch reflex, a joint flexion reflex, \glspl{cff} and \glspl{cpg}---can be seen as embedding predictable temporal regularities between observed and controlled variables within neuromuscular and skeletal systems for the purposes of effective control. These regularities are predictable because they arise as a consequence of the physical characteristics of the motor plant and the environment, which are stable over evolutionary timescales. Thus, they can be “hardwired” into neural circuitry. Their rapid, autonomous functionality can then be leveraged by supraspinal circuits, such as those in the brainstem, to guide point-to-point or rhythmic movements. The relatively short timescale over which \gls{spc} circuits operate, their relatively rigid functionality, and their need to output high-dimensional and continuously valued control signals places them near the origin of a three dimensional space we use to organize our thinking regarding the respective contributions of different brain systems to behavioral control (Figure \ref{fig:2}C). 

\begin{figure}[htbp]
    \centering
    \includegraphics[width=1\linewidth]{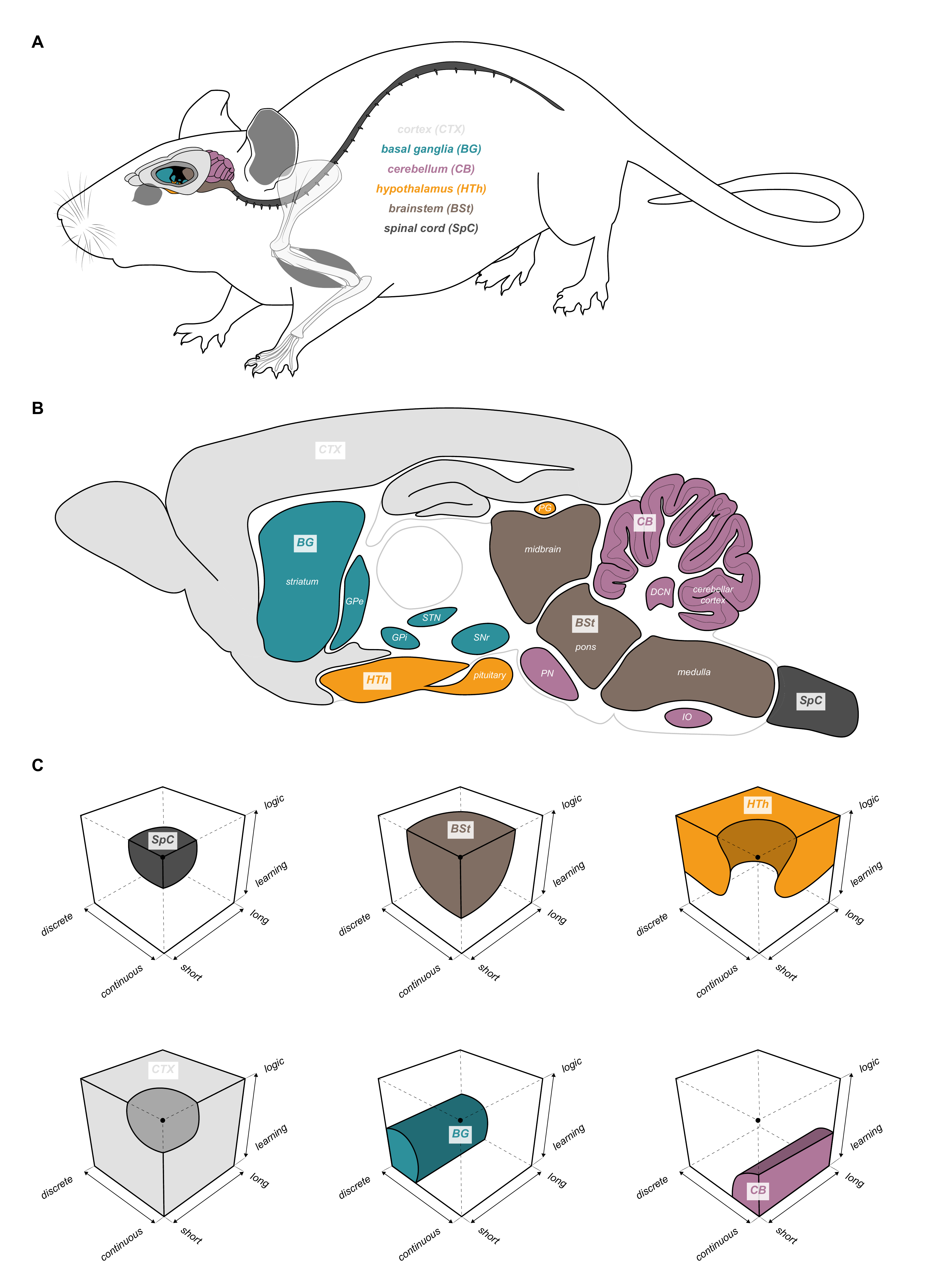}
\end{figure}
\begin{figure}[t]
    \caption{
    \textbf{Three dimensions along which brain systems differ with respect to their contributions to behavioral control.} \textbf{A)} Schematic of the rodent nervous system, highlighting six sample brain systems that we discuss in the review: spinal cord (SpC), brainstem (BSt), hypothalamus (HTh), basal ganglia (BG), cerebellum (CB), and cortex (CTX). \textbf{B)} Schematic sagittal slice of a rat brain, depicting the gross anatomical arrangement of the sampled brain systems. GPe: external globus pallidus; GPi: internal globus pallidus; STN: subthalamic nucleus; SNr: substantia nigra pars reticulata; PG: pineal gland; PN: pontine nuclei; DCN: deep cerebellar nuclei; IO: inferior olive. \textbf{C)} Focusing on the various ways in which neural systems leverage temporal regularities in the environment to support control, three characterizing axes emerge; logic to learning, short to long timescales, and continuous to discrete control. Together these define a space in which brain systems might be positioned to aid integrative thinking of how they collectively contribute to behavior. Locations of each brain system are depicted as volumes to emphasize the relative extendedness of their functionality along each dimension. The black dot denotes the origin of our proposed space.}
    \label{fig:2}
\end{figure}

\subsection{Brainstem: an interface between learning and logic systems}
\label{subsec:brainstem}
The \gls{bst} is contiguous with the \gls{spc}, sharing many of its organizational features, and can be described as containing an upper executive layer for governing somatic and autonomic reflexes to drive rhythmic or otherwise ethologically typical behaviors. It also connects directly to different sensors and effectors in the head or neck. It is divided longitudinally into regions (midbrain, pons, and medulla oblongata) and serves as a critical conduit between the cerebrum, cerebellum, and \gls{spc}. Aside from the \gls{spc}, the \gls{bst} is the only other site where lower \glspl{mn}\footnote{Motor neurons located in the spinal cord and brainstem that directly innervate skeletal muscles (includes alpha and gamma motor neurons).} originate. 

However, the \gls{bst} has distinct roles beyond those of the \gls{spc}. It interfaces with spinal reflex arcs through upper \glspl{mn}\footnote{Projection neurons located in the motor cortex and brainstem that provide mono- and/or polysynaptic inputs to lower motor neurons.}, and four of the five major descending spinal tracts\footnote{The corticospinal, vestibulospinal, tectospinal, rubrospinal and reticulospinal tracts transmit motor and postural commands from the brain to the body.} originate there. It is organized into discrete nuclei interwoven with a reticular formation that together integrate multimodal, ethologically relevant sensory inputs. Though we do not extensively cover their function here, the \gls{bst} uniquely harbors some of the primary neuromodulatory centers, including locus coeruleus; \gls{vta} and \gls{snc}; raphe nuclei; pedunculopontine and laterodorsal tegmental nuclei, which respectively synthesize and release norepinephrine, \gls{da}, serotonin and acetylcholine across a large extent of the brain, modulating behavioral responses with temporally extended effects on excitability and synaptic plasticity. \Gls{bst} circuits therefore gate and tune a diversity of behavioral processes.

As an example of its role in central pattern generation, the \gls{bst} can initiate/terminate, and control speed, gait, and direction of locomotion~\parencite{Leiras2022-by}. Stimulating the \gls{mlr} reliably triggers locomotion and adjusts gait. Successively faster walking, trotting and galloping are evoked at successively higher stimulation intensities in the decerebrate cat~\parencite{Orlovskii1966-oc,Shik1969-ux}, as in other vertebrate species~\parencite{Eidelberg1981-og,McClellan1988-wg,Coles1989-fi,Ryczko2013-qc}. Neurons in the \gls{mlr} project primarily to upper \glspl{mn} in the \gls{mrf}, which in turn provide descending input to spinal \glspl{cpg}~\parencite{Dautan2021-si}. More recently, chemo- and optogenetics have revealed that different subpopulations of \gls{mlr} and \gls{mrf} neurons can differentially initiate, maintain, arrest and modulate locomotion, depending on the strength, specificity and bilateral symmetry of the manipulation~\parencite{Bouvier2015-vt,Roseberry2016-yf,Capelli2017-gy,Caggiano2018-ya,Juarez_Tello2024-cg}.

The \gls{bst} modulates other essential behaviors, such as chewing, prehensile forelimb movements, and even foraging, through distinct control modules~\parencite{Ruder2019-mh,Ruder2021-ab,Ferreira-Pinto2021-if}. The fast, continuous feedback loops implemented by \glspl{cpg} at lower levels of the control hierarchy, are, to a large degree, controlled at slower timescales by \gls{bst} circuits in a feedforward, intermittent manner. 

As an example of a mid-level \gls{bst} controller, the \gls{sc}, in contrast to the \gls{spc}, processes multimodal sensory signals through simple spatiotemporal filters applied to sensory input that integrate evolutionarily stable dynamics of salient, behaviorally relevant stimuli, coupling them to reliably appropriate behavioral responses such as orientation, approach and evasive action. However, through interactions with higher systems, it also supports the ability to map relatively complex multimodal information to such behaviors, such as directing escape behavior towards a remembered shelter location~\parencite{Campagner2023-zy}, and even gating non-spatial behavior based on learned categorization~\parencite{Peysakhovich2024-hr}. 
	
Overall, the \gls{bst} (Figure \ref{fig:2}A-C) aggregates individual reflexes in the \gls{spc} into directed movement sequences and provides a mixed control space to modulate them both continuously (speed, direction) or discretely (gait). It can combine multiple sources of sensory information (e.g. \gls{sc}) and spatiotemporal scales over which control functions grow substantially, allowing autonomous control circuits for walking, chewing, or swallowing. Decerebrate animals can spontaneously perform these and other behaviors~\parencite{Whelan1996-pg}. However, these movements are stereotyped and “machine-like”, and by themselves \gls{bst} and \gls{spc} are limited in their ability to flexibly adapt to contextual changes~\parencite{Hinsey1930-me,Grillner1985-lz}. We thus place the \gls{bst} midway along the discrete to continuous dimension of our space, processing information over marginally longer timescales, and yet at a similar level of rigidity in its operations relative to the \gls{spc} (Figure \ref{fig:2}C). 

\subsection{Hypothalamus: setting high level control objectives}
\label{subsec:hypothalamus}
Ultimately, the control of effector systems is important for animals insofar as it promotes survival and reproduction. At any given moment, the actions that best support those superordinate goals depend on a multitude of environmental factors, from internal state variables related to nutrition, hydration, sleep and sexual maturity, to external variables that signal the time of day, safety, and presence of food, water or mates. Such factors organize various subsystems to orient actions towards the satisfaction of base requisites for survival. In vertebrates, the system generally responsible for oversight over such motivational states is the \gls{hth} (e.g., satiety and feeding centers in the ventromedial and lateral \gls{hth}, respectively~\parencite{Margules1962-cx,Hoebel1969-qd}. Here too we see that the \gls{hth} has absorbed temporal regularities from its environment.

For example, the Earth’s rotation about its own axis and revolution around the Sun delineate days and years, imposing multiscale predictability on countless vital environmental variables. Animals leverage these periodicities via anticipatory circadian and circannual rhythms that enable efficient resource allocation and appropriate sequencing of incompatible processes~\parencite{Moore-Ede1986-qt,Panda2016-aa}. While genetically determined wiring rules allow hardcoding of stable dynamics~\parencite{Zador2019-yc}, in the case of cell-autonomous circadian rhythms, genes comprise the dynamical system itself. In endogenous \glspl{ttfl}, the transcription of specialized “clock genes” is repressed by the products of their translation with a period of $\sim24$ hours~\parencite{Takahashi2017-ps}. In vitro, and in the absence of \textit{zeitgebers}\footnote{External cues that entrain autonomous biological rhythms, or “time-givers”.}, these oscillations persist. Privately they each maintain a stable degree of accuracy and precision in their period, but their free-running state inevitably leads to arrhythmicity at the ensemble level~\parencite{Welsh2004-nj}. In vivo, these distributed timekeeping mechanisms are synchronized by the hypothalamus and entrained by its chief \textit{zeitgeber}---light. 

This photic entrainment requires a subset of intrinsically photosensitive retinal ganglion cells innervating the \gls{scn} of the hypothalamus~\parencite{Berson2002-zv} that are independent of other visual information processing pathways~\parencite{Johnson1988-lu}. Critically, these \gls{scn}-projecting ganglion cells integrate luminance over long time scales, showing little to no adaptation to tonic stimuli and responding negligibly to transient ones, i.e., they embed in their response kinetics the dynamics of the exogenous day-night cycle~\parencite{Berson2002-zv}, a hallmark of timescale-matched, logic-like control as we have defined it. 

Neurons in the \gls{scn} make up one of the central pacemakers of the circadian timing system~\parencite{Welsh2010-ua}. \Gls{scn} lesions abolish the temporal organization of free-running states such as sleep, feeding, drinking, temperature and metabolic regimes, without affecting the overall distribution of sub-states visited~\parencite{Moore1997-xk}. Furthermore, \gls{scn} activity displays a circadian rhythm even when isolated from the rest of the brain, both in vitro~\parencite{Shibata1988-qi} and in vivo~\parencite{Inouye1979-wi}. Perhaps most compellingly, rhythmicity in \gls{scn}-lesioned animals is rescued by the grafting of neonatal \gls{scn} tissue, with a restored free-running rhythm matching that of the donor~\parencite{Ralph1990-ev}. Altogether, the self-perpetuating and reproducible patterns of activity within the \gls{scn} are reminiscent of the molecular \glspl{ttfl} present in every cell, with both systems baking a core aspect of the environment into their intrinsic dynamics, but still relying on regular resets to maintain global synchrony: for the \gls{scn}, those come in directly from the eye; for peripheral \glspl{ttfl}, they are inherited from the \gls{scn}, through either neural or humoral information channels.

Entrainment by other cues occurs as well, especially under conditions of constant illumination. Food availability, if confined to a small time window every day, produces a reliable rhythm in anticipatory behavior~\parencite{Mistlberger1994-tr}, which peaks around previously scheduled times long after food delivery ceases~\parencite{Boulos1980-em}. This behavioral anticipation is preceded by physiology---the activity of enzymes regulating blood glucose levels rises in advance of the predicted meal to prepare for digestion~\parencite{Moore-Ede1986-qt}.

However, food-entrained anticipatory responses do not occur for feeding intervals that deviate from the typical range of photoperiods in the wild, indicating limits on flexibility of the response. Even if rats are exposed to unnaturally short or long “days”, they fail to predict scheduled “daily” access to food~\parencite{Bolles1965-jc}. We view this rigidity in circadian timing as yet another instance where environmental stability over evolutionary time has been incorporated into properties of the system, which stands in stark contrast to the flexibility that the very same species exhibits when timing in the seconds-to-minutes range. There, because most events and actions can not be naturally synchronized to exogenous periodicities, interval timing mechanisms must be able to start and stop arbitrarily. Accordingly, the notion that there would be strict temporal limits or critical windows during which learning in this time range can occur has long been refuted~\parencite{Gibbon1977-wn}. Instead, interval timing is remarkably time-scale invariant, with the pairing of predictive cues to outcomes depending not on the absolute time delays between them (which can be arbitrarily long), but on temporal contingency~\parencite{Gallistel2000-ct}.

Further support for splitting circadian and interval timing along the logic-learning divide comes from current understanding of their respective neural substrates. In separate experiments, both \gls{scn}-lesioned mice and those with a point mutation in a clock gene that results in impaired circadian timing showed no reliable deficits when anticipating food rewards delayed by 10 seconds~\parencite{Lewis2003-oi,Cordes2008-ee}. And conversely, both decorticated rats and those bearing large bilateral lesions in either the hippocampus or the striatum, two areas strongly implicated in interval timing, showed no impairment in their ability to predict a circadian feeding opportunity~\parencite{Mistlberger1992-wc}. Thus, manipulations targeting brain regions and/or molecular mechanisms implicated in interval timing seem to produce no effects on circadian timing and vice-versa.

We thus define the \gls{hth} as more of a logic system that seeks to control high level objectives, appropriately anchoring behavioral control within its internal and environmental context. It possesses both discrete (e.g., sleep stage, wake) and continuous (degrees of hunger) control capabilities, and operates over a wide range of timescales, from months to seconds (Figure \ref{fig:2}C).

\section{A system of systems for learning to control behavior in time}
\label{sec:learning-systems}
While automatized mechanisms form sophisticated systems for controlling behavior, they are ill-suited for ingenuity. What we might call inventive behaviors originate in a collection of brain systems spanning the hindbrain to the forebrain where activity-dependent long-term synaptic plasticity significantly influences circuit function. These brain systems include the \gls{cb} in the hindbrain, and forebrain structures including the allo- and neocortices (\glsunset{ctx}\gls{ctx}), and the \gls{bg}. Although other systems utilize activity-dependent plasticity for development, injury response, or experience-based modifications~\parencite{Constantine-Paton1978-xo,Cline1990-ft,Bunday2013-iz,Rubio2020-hd}, the \gls{cb}, \gls{ctx}, and \gls{bg} are particularly capable of flexibly learning new control programs throughout adulthood, reflected in higher levels of NMDA receptors~\parencite{Monyer1994-lx} important for synaptic plasticity, compared to hypothalamic, brainstem and spinal circuits.

Again, we consider the functional role and unique contribution of these learning systems and their interactions through the lens of temporal processing and of the dynamics they control. In particular, learning systems integrate experience into some form of internal predictive model (weak anticipation)~\parencite{Dubois2003-tm,Stepp2010-qi} taking into account the idiosyncracies of experience that are not stable enough across generations so as to be absorbed by strong-anticipatory logic systems through the process of natural selection. Below we present evidence in support of a view with \gls{ctx} as a hierarchical representation learning system across the sensory-motor divide, and \gls{bg} and \gls{cb} learning multiscale policies suitable for diverse task demands. Specifically, we explore the hypothesis of \gls{bg} and \gls{cb} as hybrid hierarchical \gls{rl} systems, with distinct algorithmic approaches suited for discrete and continuous action spaces, respectively.

\gls{rl} as a framework refers to a class of behavioral problems where artificial agents or animals must learn how to obtain rewards through interaction with their environment~\parencite{Sutton2018-pf}. Within this framework, policies are learned to maximize expected future reward, by mapping situations an agent encounters in the environment (states) onto particular behaviors available in those situations (actions). Classically, \gls{rl} algorithms were studied in simplified scenarios where state variables, actions, and rewards were pre-defined and directly available to agents. For animals, however, the significance of high dimensional sensory and motor signals must often be learned through experience, potentially involving long spatiotemporal integration to either estimate the state of the world, or to sequence and shape more granular behaviors. This involves (cortical) representation learning~\parencite{Radulescu2021-bj}, a process that is likely to require many sampled observations from the environment and that is thus relatively slow. 
 
In the following sections we illustrate how anatomical differences between \gls{ctx}, \gls{bg} and \gls{cb} make them ideally suited for implementing components of algorithmic approaches to solving \gls{rl} problems presented to animals.

\subsection{Cortex: a representation learning system for hierarchies of states, rewards and actions}
\label{subsec:cortex}
Cortex has long been viewed as a hierarchical, feedforward processing system, and indeed many aspects of sensory cortical neural activity can be explained through this lens. For example, deep neural networks trained on unsupervised or task objectives resemble cortical representations at various stages and spatial scales~\parencite{Atick1990-uy,Karklin2009-vq,Yamins2016-ag,Cowley2023-ec}. In machine learning, deep \gls{rl} combines representation learning and reward-guided policy learning, allowing simulated agents to reach human-level or even super-human performance on complex tasks~\parencite{Mnih2015-mc,Dabney2018-qr}. However, these feedforward architectures are limited in their ability to integrate information over time. In contrast, \glspl{rnn} maintain information about past inputs in their activation state, thereby using internal network dynamics for more effective policy learning~\parencite{Han2020-of}. Recurrent connections are an essential feature of cortical networks and express various forms of short- and long-term plasticity that enable the encoding and learning of temporal patterns~\parencite{Buonomano1995-lw,Goel2016-kc}. Slow intrinsic timescales can arise from clustered connectivity~\parencite{Litwin-Kumar2012-el} and be extrinsically modulated by internal states such as attention or arousal~\parencite{Huang2019-gr,Zeraati2023-ox}. Such features may provide \gls{ctx} with a temporal basis for predicting environmental states or actions that mirror the increasing spatial scale along its representational hierarchy~\parencite{Born2005-wq,Rust2010-sb,Murray2014-ul}. 

Recurrent connectivity is prominent in the stereotypical local and long-range circuit architecture of \gls{ctx}~\parencite{Douglas2004-pr,Bastos2012-cw}. Feedforward inputs originate in upper layers and excite layer IV pyramidal neurons and inhibitory interneurons, which are critical for maintaining balance between excitation and inhibition~\parencite{Tsodyks1997-mj,Troyer1998-iy,Taub2013-bd}. Excitatory connectivity between layers II/III and layer V provides strong within-region recurrence, a motif that increases along the cortical hierarchy and is suggested to underlie longer processing timescales~\parencite{Hasson2015-uq} for evidence accumulation~\parencite{Cavanagh2020-ko}. Deep layers send feedback projections to superficial layers within \gls{ctx}, as well as extracortical projections to the \gls{bg}, thalamus, \gls{cb}, \gls{bst} and \gls{spc}. In addition to providing major sensory drive to layer IV cortical neurons (for most modalities), the thalamus forms reciprocal connections with various cortical areas, creating recurrent loops for slower temporal processing that integrate information across areas, and allow for influence of contextual priors and attentional modulation~\parencite{Harris2019-cy,Zajzon2019-qg,Kumar2022-na}. These mesoscopic connectivity patterns add to \gls{ctx}’s ability to use recurrent dynamics to integrate information over larger extents of time~\parencite{Goldman2009-yl,Mante2013-ur}. 

The aforementioned circuit properties seem ideal for allowing \gls{ctx} to hold onto past observations, but to what end? Canonical cortical computations are thought to involve predictive processing~\parencite{Wacongne2011-go,Keller2018-yu}, where generative models learn predictions that are passed across hierarchical levels~\parencite{Ororbia2022-om}. Hypotheses include the propagation of state and action prediction errors~\parencite{Rao1999-ii,Friston2010-gr,Rao2024-nc}, or state prediction along the hierarchy~\parencite{Heeger2017-qf}. Such predictions might be facilitated through the straightening of dynamic information within neural space, allowing accurate estimation of future states through simple linear extrapolation~\parencite{Henaff2019-lp,Henaff2021-rh}. Interestingly, a series of hierarchically arranged cortical regions implementing such operations naturally gives rise to a hierarchy of timescales~\parencite{Murray2014-ul}, and indeed, timescales of information integration span tens to hundreds of milliseconds in early sensory areas to minutes or longer in higher-order regions~\parencite{Hasson2015-uq,Siegle2021-xq,Brunec2022-rl}, facilitating longer contextual priors~\parencite{Heeger2017-qf} and increasing temporal receptive windows. These features of higher \gls{ctx} can be seen as supporting multimodal, context-dependent, temporally-extended abstractions to inform more temporally extended aspects of behavior.

 A similar picture of temporal hierarchy emerges in motor areas. Classic experiments of Fritsch, Hitzig and Ferrier, wherein brief electrical stimulation at different sites in \gls{m1} produced muscle twitches in different parts of the body, revealed a somatotopic motor map~\parencite{Ferrier1873-nx,Fritsch2009-hm}. However, long electrical stimulation trains (e.g., 500 ms) can produce coordinated, goal-directed movements involving multiple body parts~\parencite{Graziano2005-ey}, and certain subregions seem associated with movement categories~\parencite{Graziano2010-zf}. Stimulating one \gls{m1} area can trigger defensive forelimb and facial movements, while another elicits hand-to-mouth actions. In rodents, different \gls{m1} sites map to exploratory or defensive whisker movements~\parencite{Haiss2005-nq}. Thus, movement representations, even at low levels in the cortical hierarchy, reflect behavioral categories, providing a discrete set of selectable motor programs, perhaps learned so as to achieve specific classes of behavioral objectives.

With the rise of large-scale cortical population recordings, emphasis is increasingly placed on dynamic trajectories of neural activity for movement preparation, execution and timing~\parencite{Shenoy2013-yv,Gallego-2017,Gamez2019-gt,Wang2022-ho}. Again, we see evidence of hierarchical action representations at increasing levels of abstraction. \Gls{m1} populations display distinct representations of forward vs. reverse forelimb rotations that are low-dimensional and untangled compared to muscle representations, and form discrete categories~\parencite{Russo2018-zx}. However, while repeated rotational movements trace the same trajectory in \gls{m1} neural space, a sequence of rotations is accompanied by a helical trajectory in the \gls{sma}, a structure that encodes the temporal location within the motor sequence~\parencite{Russo2020-ej}. Both of these representations allow predictions at different scales, within individual movement cycles in \gls{m1} and within a sequence of movement cycles in \gls{sma}, consistent with the idea that higher regions in the motor hierarchy predict the future states of activity in lower regions~\parencite{Shipp2013-bv}. Hierarchical temporal processing therefore appears to represent a fundamental organizing principle of cortical representations both in sensory processing and motor execution~\parencite{Dum2002-cz}, leading to increasing spatiotemporal scales of representation and greater degrees of abstraction at successive levels. \Gls{ctx} thus occupies arguably the largest volume in our proposed space, occupying all but the region closest to the origin where the most rigid, rapid, and continuous neural mechanisms for control reside (Figure \ref{fig:2}C). 

Interestingly, the full stack of \gls{ctx} representations are fed to the \gls{bg} and \gls{cb}, two other learning systems that we propose to use \gls{ctx} representations for learning sets of hierarchically organized control policies. 

\subsection{Basal ganglia: a policy learning system for discrete control}
\label{subsec:basal-ganglia}
The \gls{bg} are an evolutionarily ancient collection of nuclei spanning the forebrain and midbrain, implicated in procedural learning and motor function~\parencite{Grillner2016-un}. They receive excitatory input from diverse state-action representations in cortex, thalamus, amygdala and hypothalamus, funneling information from the striatum through successive processing stages, ending in the \gls{gpi}\footnote{Due to its embedding in the fiber tract of the internal capsule that passes through the cerebral peduncle, the rodent homolog to the primate \gls{gpi} has historically been termed the endopeduncular nucleus, but here we use \gls{gpi} to refer to both.} and the \gls{snr}. From here, projections to \gls{bst} and thalamic nuclei~\parencite{McElvain2021-xh} enable the \gls{bg} to influence motor or other processes either directly or indirectly through modulation of thalamo-cortical function. 

The striatum receives extremely dense input from midbrain \gls{da} neurons~\parencite{Bertler1959-qc}, which innervate large volumes of striatal tissue~\parencite{Matsuda2009-lz}. The phasic activity of midbrain \gls{da} neurons bears striking resemblance to a teaching signal within computational models of \gls{rl}, the \gls{td} \gls{rpe}, that updates representations of value that encode expected future reward~\parencite{Montague1996-gm,Schultz1997-cm}. Value estimates play a crucial role in a class of \gls{rl} models, where they are used to derive policies that specify which actions to take depending on the current state of the world~\parencite{Sutton2018-pf}. The convergence of \gls{da} teaching signals, with cortical sensory and motor information is thought to enable learned policies for behavioral control through synaptic plasticity at cortico-striatal synapses~\parencite{Reynolds2001-zu,Shen2008-hs,Iino2020-st}. In line with this, direct stimulation of \gls{da} populations has long been known to reinforce behavior~\parencite{Olds1954-nq,Tsai2009-ke}, providing cohesive support for the idea that the \gls{bg} act as a policy learning system.
 
The \gls{bg} are also characterized by a parallel circuit architecture. Information from distinct cortical and thalamic regions is carried along relatively segregated channels that, broadly speaking, reflect the hierarchical organization of \gls{ctx}. Inputs from similar levels of the cortical hierarchy, but originating from both sensory and motor domains, largely converge in particular regions of the striatum~\parencite{Hunnicutt2016-cp,Hintiryan2016-da,Foster2021-bw}. These zones of convergence define the starting points for a set of parallel circuits that are classically split into sensorimotor, associative, cognitive, and limbic according to the information content they inherit~\parencite{Alexander1990-rh}. Importantly these circuits also inherit the varying degrees of temporal extendedness of the cortical hierarchy. This suggests that any policy learning algorithm implemented by \gls{bg} circuitry is applied to diverse state and action information in parallel, informing behavior over different temporal horizons. However, the functional significance of this parallelism, and how \gls{bg} circuitry operates on these state and action spaces to influence behavior are debated~\parencite{Bornstein2011-nl,Klaus2019-ne,Markowitz2020-ye,Dhawale2021-kr,Monteiro2023-cg}. 

Ever since \gls{rl}-like algorithms were first associated with the \gls{bg}, it has been appreciated that such computations require access to signals that encode the what and when of past observations~\parencite{Schultz1997-cm}. In other words, \gls{rl} requires a temporal basis for encoding the dynamic value of current state-action combinations. Perhaps not surprisingly then, disruption of \gls{bg} function can alter timing behavior in humans and non-human animal models across a range of tasks, and functional imaging and neural recordings reliably identify signals in the \gls{bg} that covary with timing behavior~\parencite{Merchant2024-je}. An emerging theme across many systems is that the timecourse with which neural activity evolves along a trajectory in neural population state space may provide the timebase for the computations a population is performing~\parencite{Mauk2004-jl,Wang2017-xu,Paton2018-nn}. In striatal circuits this provides a plausible mechanism for approximating the types of time-varying functions required for value-based \gls{rl}~\parencite{Jin2009-yk,Gouvea2015-me,Mello2015-ny,Kim2018-yr,Rodrigues2023-ic}. However, a debate has emerged regarding whether time-varying patterns of activity are used to control learned kinematic features~\parencite{Rueda-Orozco2015-gz,Park2020-bz,Dhawale2021-kr,Mizes2023-ut}, or abstract, discrete transitions between segments of behavior~\parencite{Markowitz2020-ye,Monteiro2023-cg}. 

Systematically perturbing the timecourse of moment-by-moment activity patterns, while leaving their structure intact, represents an incisive means of probing how temporal evolution of population activity is used to guide behavior~\parencite{Banerjee2021-cb}. Recent studies experimentally manipulated the temperature of neural tissue to test hypotheses regarding the behavioral role of specific circuits. For example, the warming and cooling of area HVC (proper name) in songbirds, where neurons fire at precise moments within the bird’s song, contracted and stretched the timecourse of the song, while leaving its spectral properties and overall syllabic structure unchanged~\parencite{Long2008-hd}. Similarly, in neotropical singing mice, cooling of the orofacial motor cortex strongly affected song timing by monotonically increasing the overall song duration, despite no effects on the duration of individual notes~\parencite{Okobi2019-cs}. And lastly, manipulating the temperature of nearby regions in the speech motor cortex of humans produced rescaling of speech in time, or a degradation in speech quality, depending on the targeted site, revealing a functional architecture within the motor cortex with respect to different aspects of speech~\parencite{Long2016-ke}. 

Inspired by this body of work, we have recently manipulated the temperature of striatal tissue in rats in the context of an interval categorization task involving both discrete categorical judgments and their report via continuous execution of lateralized whole-body movements~\parencite{Gouvea2014-go}. Briefly, slowing and speeding the moment-by-moment evolution of striatal population activity via cooling and warming biased animals towards short and long judgments, respectively, thus providing strong evidence for striatal population speed playing a causal role in timing. However, we did not observe similar effects of temperature on the kinematics of choice movements. Instead movement was slowed by both warming and cooling the striatum. This effect resembled a distinct, non-monotonic effect of temperature where both cooling and warming lowered baseline firing rate~\parencite{Monteiro2023-cg}. Altogether, a parsimonious interpretation is that the \gls{bg} influence the timing of discrete transitions between actions through temporal scaling of higher-dimensional striatal population firing. Their main contribution to continuous control appears to be through a gain-like signal controlling movement vigor, encoded in the overall activity of direct-pathway medium spiny neurons~\parencite{Panigrahi2015-gp,Cruz2022-um}. Intriguingly, while \gls{da} tone is clearly capable of positively modulating the gain of movements through vigor~\parencite{Turner2010-mg,Beierholm2013-zs}, its effects on the temporal placement of transitions appears to be distinct~\parencite{Soares2016-wp,Motiwala2022-fe}. Whether this distinction arises from dissociable effects of \gls{da} on overall striatal activity levels versus their temporal evolution remains to be seen. 

Why employ different mechanisms for discrete and continuous control? Continuous online control can adjust precisely to fluctuations, however, continuous systems often possess strongly nonlinear dynamics where small input changes lead to large output changes, leading to highly diverse movement trajectories that nonetheless succeed in achieving their goal~\parencite{Graziano2006-rv,Li2016-yg}. In the face of such challenges, reliable performance may benefit from multiple stable modes that structure high-dimensional dynamics around attractor states~\parencite{Huber2002-ki}, effectively creating a discretization of action space. Division into discrete alternatives allows for selection, for instance to guide movements towards one goal or another, while leaving the continuous trajectory between modes flexible. This discrete space improves sample efficiency by limiting learning to stable action alternatives that can be explored quickly~\parencite{Kober2013-fm}, and discrete control is provably optimal under time pressure~\parencite{Lasalle1959-ky}. It also reduces the need for compensatory actions and allows a discretely chosen action to be executed until convergence in continuous dynamics~\parencite{Huber2002-ki}. \Gls{bg} outputs have been proposed to interface with discrete action spaces by switching dynamical regimes in other regions, such as in motor cortex or the \gls{bst}, that encode learned or hardwired motor programs, respectively~\parencite{Roseberry2016-yf,Logiaco2021-ds,Arber2022-ck}. 

The above suggests that \gls{bg} circuits learn to promote those actions that are associated with better outcomes but that the moment-by-moment execution of the movements that correspond to those actions is handled by circuits elsewhere. We thus place the \gls{bg} in our space as a learning system, operating on a range of timescales, but largely on discretized control spaces (Figure \ref{fig:2}C). This, however, begs the question: which circuits acquire learned aspects of continuous control programs in order to properly execute actions?

\subsection{Cerebellum: a policy learning system for continuous control}
\label{subsec:cerebellum}
The \gls{cb} is a hindbrain structure implicated in continuous coordination of movement, as well as correcting for disturbances and delays in the sensorimotor system by purportedly learning internal models that predict the sensory consequences of motor output~\parencite{Wolpert1998-st,Bastian2006-sw}. It shares several features with the \gls{bg}. The \gls{cb} processes diverse information from the cortical hierarchy~\parencite{King2023-px} and sends outputs directly to motor control structures in the \gls{bst}, as well as ascending projections to the thalamus for indirect control of cortical activity through thalamo-cortical circuitry. The \gls{cb} has also been implicated in timing function, albeit on the shorter timescale of tens to hundreds of milliseconds as compared to the seconds to minutes timescales most often associated with the \gls{bg}~\parencite{Ivry1989-bp,Merchant2024-je}. The \gls{cb} receives sensory information at short latency, due to its proximity and connection to brainstem and spinal sensory circuits. Such inputs are thought to subserve more rapid and continuous involvement of \gls{cb} circuits in modulating ongoing behavior in response to observed sensory signals. 

The \gls{cb}’s highly structured architecture is thought to enable precise temporal processing. \Gls{mf} inputs carry contextual information originating in \gls{ctx}, \gls{bst}, or \gls{spc}, to the cerebellar cortex, where they synapse onto \glspl{gc} and other \gls{cb} interneurons~\parencite{Marr1969-hf,Albus1971-ud,D-Angelo2013-eo}. \Glspl{gc} drive inhibitory feedback onto themselves via Golgi cells, a feature thought to contribute to the network transforming relatively slower \gls{mf} signals into a rich variety of high-dimensional, rapidly fluctuating \gls{gc} activity profiles~\parencite{Mauk2004-jl,Garcia-Garcia2024-ch}. These temporal patterns represent a basis expansion of \gls{mf} inputs~\parencite{Lanore2021-jh}, and plasticity at \gls{gc}-Purkinje cell (\glsunset{pc}\gls{pc}) synapses driven by error signals carried by \glspl{cf} is thought to allow \glspl{pc} to learn complex, time-varying signals~\parencite{Warren2016-iq} essential for motor and cognitive control, which ultimately exert their influence via projections to the deep cerebellar nuclei.

Several influential models have been developed to explain the computational principles of learning and control in this circuit. The classic Marr and Albus theories emphasized a capacity for learned pattern recognition and classification due to the massive dimensionality expansion of \gls{mf} input within \glspl{gc}~\parencite{Marr1969-hf,Albus1971-ud}, but underemphasized the role of dynamics and continuous control functions. Ito and colleagues incorporated a temporal dimension and described the circuit as learning to minimize control errors with internal models that predict the sensory consequences of movement~\parencite{Ito1972-bh}. This suggests a forward model that predicts future states given the current state and ongoing actions of the nervous system. If such a forward model is included within a control system, it can be leveraged to produce appropriate control output. Specifically, if a desired target state is provided as an objective, the forward model can compute the expected future state of the system given current conditions, which can then be compared to the target. If there is an error between the expected and desired future state, the system can adjust its output to minimize that error. However, this would involve interaction with an inverse model to resolve the distal error problem that arises from the fact that sensory errors and motor commands occupy different coordinate frames~\parencite{Jordan1992-ey}. The accuracy of the forward model could in principle be maintained by continually comparing the predictions from previous time points to current observations and using any errors to update the parameters of the forward model in a supervised manner. 

The view of the \gls{cb} as learning a forward model for control suggests a combination of motor commands and sensory and motor errors in the \gls{cb} circuitry. Indeed, \gls{cf} activity can encode predicted reach location, or reach errors relative to a target at different points in time~\parencite{Kitazawa1998-hd}. However, studies have also revealed significant reward-related signaling, in humans, non-human primates, and mice, and even \glspl{rpe} that resemble the kinds of \gls{td} \glspl{rpe} traditionally associated with the \gls{da} system in the \gls{bg}~\parencite{ODoherty2003-sn,Ohmae2015-zx,Heffley2018-au,Kostadinov2019-vr,Sendhilnathan2021-wt}. Such observations suggest that the supervised learning view may need to be expanded to include a role for reinforcement and reward~\parencite{Hull2020-di}. Interestingly, experiments involving visuomotor adaptation to mirror-reflected feedback suggest that sensorimotor adaptation, widely thought to reflect updating of forward models, is instead the product of a \gls{rl}-like process to directly update a continuous control policy~\parencite{Hadjiosif2021-ox}. 

Taking into account anatomical and functional data alike, there is growing evidence that the \gls{cb} might contribute more than learning forward models~\parencite{Ito2008-pr}. Instead, its role in movement coordination, reward-related signaling, and potential for implicit adaptation aligns it with a \gls{rl} system for continuous control policies, where reward could be parametrized as the difference between intended and observed action outcomes. The \gls{cb}’s high dimensionality supports complex continuous control functions, and gradient-based policy learning methods may enhance \gls{cb}'s role in optimizing control policies compared to value-based \gls{rl} methods typically linked with \gls{bg} circuits~\parencite{Kober2013-fm}. For these reasons, we place \gls{cb} in our space as a learning system concerned with more continuous forms of control, both in terms of the nature of the action spaces on which it operates and in time, drawing it naturally toward operation at higher time resolution and shorter timescales (Figure \ref{fig:2}C).

\subsection{A parting example for thinking about layered, multisystem control of behavior}
\label{subsec:parting-eg}
The diverse control mechanisms in the \gls{spc}, \gls{bst}, \gls{ctx}, \gls{bg}, and \gls{cb} usually work seamlessly together, complicating efforts to isolate their unique contributions to behavior. However, visuomotor adaptation of arm movements is instructive in highlighting division of labor between brain systems. Classically, subjects reach to a target with their arm occluded but with a cursor representing their hand position. When perturbations are applied to create visuomotor errors, subjects adapt in ways that reveal surprising signatures of multiplicitous control. For example, if a 45-degree rotation is applied to the cursor relative to hand position, subjects initially miss the target but gradually adapt to be more accurate. Removing the rotation reveals after-effects, reflecting learning that corrects for the rotation. When subjects were explicitly instructed to aim 45 degrees from the target, they immediately corrected for the rotation. Surprisingly, they then drifted away from cursor accuracy, suggesting an implicit learning process driven by sensorimotor errors~\parencite{Mazzoni2006-fd}. More bizarre still, over time this error peaked and again diminished, suggesting another correction process. Two learning processes, one driven by arm-cursor errors and another by cursor-target errors seem to vie for control under the command of explicit instruction. Might these two learning processes map to particular aforementioned learning systems, and if so, how are we to understand their roles in controlling behavior more formally?

Visuomotor mirror reflection experiments suggest that implicit adaptation reflects not forward model adjustments but direct policy updating~\parencite{Hadjiosif2021-ox}, in line with our proposal that \gls{cb} functions as a \gls{rl} system for continuous control. Reduced drift in patients with \gls{cb} dysfunction following application of an explicit pointing strategy in the face of an applied rotation may similarly reflect deficits in updating an implicit, continuous control policy, while those with prefrontal cortex damage display intact and rapid implicit adaptation alongside incomplete use of the explicit strategy~\parencite{Taylor2014-uu}. Likewise, patients with Parkinson’s disease, who lack sufficient \gls{da} tone in sensorimotor circuits of the \gls{bg}, also display signs of impairment in explicit aiming~\parencite{Tsay2022-kr}. Thus, effective use and/or updating of an explicit strategy to correct for performance errors appears to involve a combination of the prefrontal \gls{ctx} and the \gls{bg}, consistent with our proposal. 

\section{Conclusion/summary}
\label{sec:conclusion}
“The discovery of simple and uniform principles, by which a great number of apparently heterogeneous phenomena are reduced to coherent and universal laws, must ever be allowed to be of considerable importance…”~\parencite{Young2019-te}. We find that viewing brains as layered~\parencite{Wilson2022-dr,Matni2024-zp}, predictive controllers of behavior reveals a centrality for temporal processing in potentially reducing a great number of heterogeneous phenomena to more universal laws. On the other hand, while we emphasize the centrality of time itself, we acknowledge that the perspectives presented here relegate time estimation per se to a somewhat subordinate level relative to the more fundamental need for essentially all neural mechanisms to absorb temporal regularities in service of effective control~\parencite{Paton2018-nn}. Indeed, consider the proposal that the \gls{hth}, \gls{ctx}, \gls{bg}, and \gls{cb} are responsible for timing intervals over longer to shorter scales~\parencite{Buhusi2005-ws}. There is no superordinate organizing principle derived from the consideration of time estimation as a fundamental function of neural systems that explains such distribution of labor. However, division of timescales may be seen to reflect the nature of temporal structure that must be absorbed into these systems to support their contributions to behavioral control, namely to set high level objectives for control in the case of \gls{hth}, to learn hierarchies of state and action representations in the case of \gls{ctx}, and to learn policies that map such representations to discrete and continuous action spaces, in the case of \gls{bg} and \gls{cb}, respectively. Given the breadth of control problems the brain is faced with, we have included only a small, but we hope representative, sample to illustrate our perspective, in an attempt to spark a more unified and less balkanized approach towards a computational understanding of the neurobiology of behavior. 

\section{Acknowledgments}
\label{sec:acknowledgments}
We thank T. S. Duarte, S. A. Zamora, M. Sousa, T. Monteiro,, J. Krakauer, and D. McNamee for their comments on the manuscript. 
Funding: HHMI International Research Scholar Award to J.J.P. (55008745); a European Research Council Consolidator grant (DYCOCIRC - REP- 772339-1) to J.J.P. Champalimaud Foundation internal funding to F.S.R. Simons Foundation Transition to Independence Award to C.H.

\section{Author contributions}
\label{sec:contributions}
C.H., F.S.R. and J.J.P. developed the ideas and wrote the manuscript collaboratively.

\section{Declaration of interests}
\label{sec:conflict}
The authors declare no competing interests.

\newpage

\printbibliography

\printglossaries

\end{document}